\documentclass[3p,times,twocolumn]{elsarticle}

\usepackage{ecrc}

\volume{00}

\firstpage{1}

\journalname{Nuclear Physics B Proceedings Supplement}

\runauth{}

\jid{nuphbp}

\jnltitlelogo{Nuclear Physics B Proceedings Supplement}

\usepackage{amssymb}

\usepackage[figuresright]{rotating}

\begin{document}

\begin{frontmatter}



\dochead{}

\title{Search for Dark Matter at CMS}


\author{Steven Lowette\fnref{lowette}}
\ead{steven.lowette@cern.ch}
\fntext[lowette]{on behalf of the CMS Collaboration}

\address{Vrije Universiteit Brussel, Pleinlaan 2, 1050 Brussel, Belgium}

\begin{abstract}
The results are presented from searches with the CMS experiment for directly-produced dark matter particles. All these searches use the full LHC Run-I dataset of $\sim 20 \, \mathrm{fb}^{-1}$ of proton--proton collisions at $8 \, \mathrm{TeV}$ centre-of-mass energy. Final states with a monojet, monophoton, and monolepton signature are considered, as well as processes with dark-matter particles produced in association with top quarks. Most of these results have been interpreted using an effective field theory approach, while first results are also reported with simplified models.
\end{abstract}

\begin{keyword}
CMS \sep LHC \sep Dark Matter

\end{keyword}

\end{frontmatter}


\section{Dark matter searches at colliders}
The hypothesis of an invisible form of matter in our universe, called \emph{dark matter}, is already more than 80 years old. Despite many pieces of evidence of the existence of dark matter, which have been collected so far from gravitational effects over a large span of astronomical scales, still very little is known of this elusive type of matter.

The many open questions have led over the years to a plethora of theoretical proposals for dark matter, many of which extend the standard model of particle physics to also describe a candidate dark-matter particle. On the experimental side, a large suite of experiments pursue the hunt for the first confirmed observation of a non-gravitational dark-matter interaction. These experiments can be classified in three categories: direct-detection searches, indirect-detection searches, and collider searches. Searches for direct detection of interactions between dark matter and normal matter try to establish the small effects of elastic collisions of dark matter particles in the earth's vicinity on nuclei in very sensitive underground experiments. The indirect searches look for an excess of photons, positrons, neutrinos, etc. that could result from the annihilation of pairs of dark matter particles in the galaxy, the sun, the earth, etc. The collider searches, finally, aim to establish creation of dark matter particles in the laboratory. All these search strategies bring complementarity through different sensitivities to the details of the SM--dark-matter interaction, the detector, and astrophysics assumptions.

Searches for dark matter production at colliders are in general looking for collisions where, transverse to the beamline, a momentum imbalance is created by the dark-matter particles, which escape detection because of their (quasi-)stability and very low interaction cross section with normal matter. Two production modes can be distinguished. In the first case, other heavier new states are produced first, which then subsequently decay down to the lighter stable dark matter particles at the end of a decay chain. One example is supersymmetry with $R$-parity convervation, where the lightest supersymmetric particle is necessarily stable. Another example is the production of a Higgs boson, which in case of a so-far undetected coupling of the Higgs boson to sufficiently light dark matter would result in a portion of the Higgs decays taking place invisibly.

A second production mode which can be considered is the direct production of dark-matter particles. In this case, typically a pair of dark matter particles is produced through some mediator that is ``off-shell''. The presence of other particles, like an initial state radiation (ISR) gluon or photon, is then required to recoil against the dark-matter candidates to render them detectable.

In this paper, five results are presented of searches for direct production of dark-matter particles with the CMS detector~\cite{cms} at the Large Hadron Collider (LHC) at CERN. All these searches use the full LHC Run-I dataset of $\sim 20 \, \mathrm{fb}^{-1}$ of proton--proton collisions at $8 \, \mathrm{TeV}$ centre-of-mass energy collected by CMS. The theoretical basis for these searches and their interpretations can be found in Refs~\cite{Goodman:2010yf,Bai:2010hh,Goodman:2010ku,Rajaraman:2011wf,Fox:2011pm,Bai:2012xg,Cheung:2010zf,Beltran:2010ww,Lin:2013sca,Andrea:2011ws,Agram:2013wda}.

\section{Effective operators and simplified models}

The interaction that gives rise to the dark-matter production has often been approximated using effective operators in an effective field theory (EFT) approach, which integrates out the details of the mediator. This allows to stay fairly model independent, and to restrict the parameter space to the mass of the dark-matter particle, and the EFT scale $\Lambda = M / \sqrt{g_\chi g_q}$, where $M$ is the mass of the mediator, and $g_\chi$ and $g_q$ are the couplings of the mediator to the dark matter and quarks, respectively. In such an EFT context, it becomes simple to translate from the collider setting to the picture of dark matter scattering off nuclei, allowing for a comparison of the results of collider searches with the indirect-detection experiments.

In making such interpretations in an EFT model, it is important to stress the limitations of that approach. For the EFT to provide a realistic description of the hard interaction, the mediator mass needs to be (much) larger than the energy scale of each collision. In direct-detection experiments, the energy transfer is very small, and the use of an EFT is mostly well justified. At the LHC, however, the energy scales are much higher, which restricts the range of applicability to mediator masses in the TeV range or higher.

A second limitation to keep into mind, is that EFT limits, as they are usually presented, assume the interaction linking the standard model to the dark sector to be described by one operator. This is not necessarily the case, like for instance for the electroweak interaction, which has a V--A structure.

\section{Searches in monojet, monophoton and monolepton final states}

The searches presented in the monojet~\cite{cmsmonojet}, monophoton~\cite{cmsmonophoton}, and monolepton~\cite{cmsmonolepton} final states all share the feature of a particle radiating from a quark line in the initial state radiation, hence providing the necessary recoil to make the pair of produced dark-matter particles appear in the detector as missing transverse momentum, $E_{\rm T}^{\rm miss}$.

The event selection in the monojet search is driven by the used $E_{\rm T}^{\rm miss}$ trigger, which was measured to become fully efficient for events with missing energy above $250\,\mathrm{GeV}$. One central jet is required with $p_{\rm T} > 110 \, \mathrm{GeV}$, while a second softer jet is allowed in the vicinity of the hard jet. A series of dedicated jet identification requirements is imposed, which was shown to effectively suppress instrumental and other sources of fake $E_{\rm T}^{\rm miss}$. Electron, muon, and tau vetoes, finally, are put in place to reject the large background of leptonic ${\rm W}$ decays.

After applying this set of selection requirements, the remaining background is dominated by ${\rm Z} \to \nu\nu$, with a subdominant contribution from ${\rm W} \to \ell\nu$, where the lepton is not reconstructed, doesn't satisfy the identification requirements of the veto, or is invisible because it went out of detector acceptance. All other backgrounds -- top, QCD multijets, ${\rm Z}\to\ell^+\ell^-$ -- are much smaller. The two dominant background components are both estimated from data using a ${\rm Z}\to\mu^+\mu^-$ control event sample, accounting for the muon as an undetected particle as appropriate and applying the necessary correction factors.  The data was found to be in good agreement with the background expectations.

\begin{figure}[htb!]
  \begin{center}
    \includegraphics[scale=0.35]{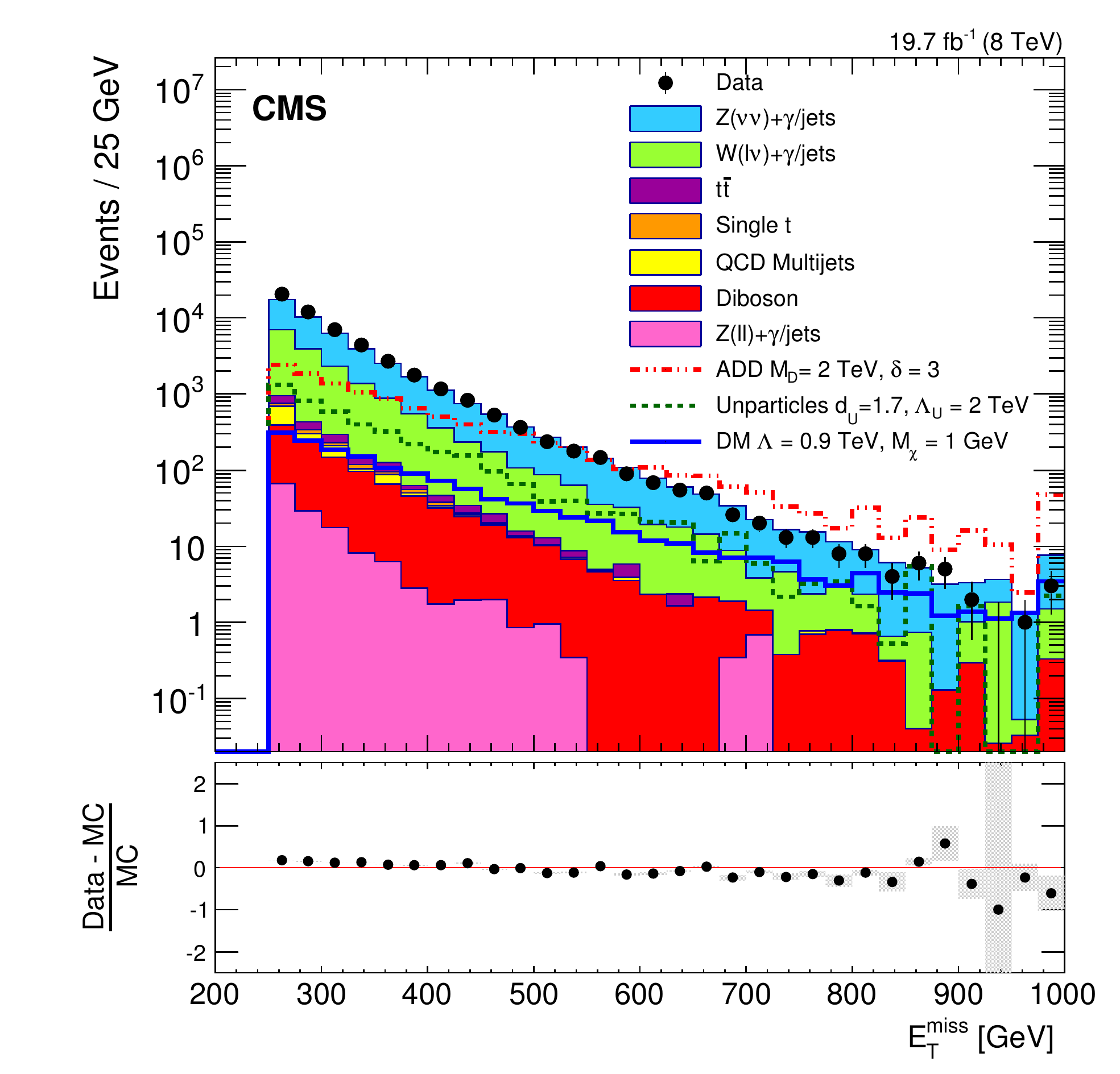}
  \end{center}
  \caption{Distribution of $E_{\rm T}^{\rm miss}$ after the monojet selection described in the text, showing data, background simulation, and signal expectation for several models of new physics leading to the monojet signature.}
  \label{fig:monojet}
\end{figure}

In Figure~\ref{fig:monojet}, the $E_{\rm T}^{\rm miss}$ distribution is shown, comparing the data with MC simulation estimates of all individual background components, where the simulation is scaled to the recorded integrated luminosity according to the individual cross sections. Expectations for signals from several different new-physics models are shown overlayed, one of these being a scenario with a dark-matter particle of $1\,{\rm GeV}$ mass.

The monophoton search shows many similarities with the monojet search described above. Also here, the event selection is driven by the trigger, in this case a single-photon trigger, which leads to the requirements $p_{\rm T} (\gamma) > 145 \, \mathrm{GeV}$ and $E_{\rm T}^{\rm miss} > 140 \, \mathrm{GeV}$. Additionally, the photon is required to be back to back in azimuth with the missing momentum vactor. The selected photon is subjected to several identification requirements, to ensure the purity in prompt photons of the sample. Also here, lepton vetoes, as well as a hadronic veto, are employed to reject ${\rm W}\gamma$ and other backgrounds.

After these selections, invisible and leptonic decays of ${\rm Z}$ and ${\rm W}$ bosons consitute again the main ${\rm Z}\gamma$ and ${\rm W}\gamma$ backgrounds. Here, additional subdominant backgrounds arise from misidentification as a prompt photon of objects like electrons from ${\rm W}\to{\rm e}\nu$ decays, jets in QCD multijet events, and energy deposits in the electromagnetic calorimeter from non-collision beam-halo particles. Backgrounds are estimated from simulation as well as from data control samples in cases where no reliable simulation is available. The data was found to be in good agreement with the background expectations.

\begin{figure}[htb!]
  \begin{center}
    \includegraphics[scale=0.35]{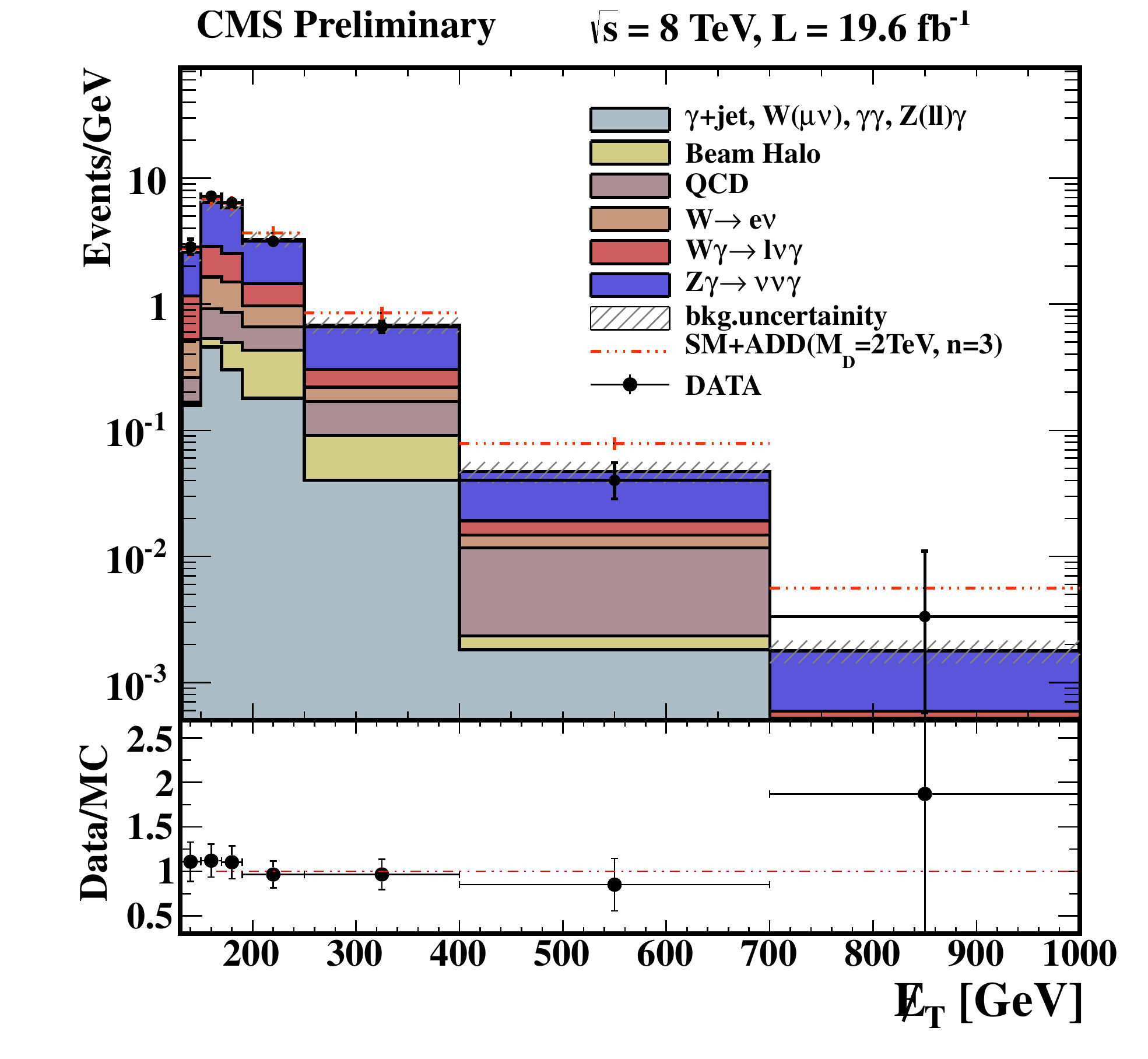}
  \end{center}
  \caption{Distribution of $E_{\rm T}^{\rm miss}$ after the monophoton selection described in the text, showing data, estimated background contributions, and signal expectation for an example model of new physics leading to the monophoton signature.}
  \label{fig:monophoton}
\end{figure}

The $E_{\rm T}^{\rm miss}$ distribution after the above monophoton selections is shown in Figure~\ref{fig:monophoton}, comparing data, the estimated background components, and the expected signal from an example new-physics model.

The monolepton search, finally, looks for events with only an electron or muon detectable, coming from a ${\rm W}$ boson, which in the case of dark-matter production would be radiated off an incoming quark. A peculiarity in this scenario is the possible interference between diagrams with identical initial and final states, but where the dark matter couples to either up quarks or down quarks. To account for this possible interference, a parameter $\xi$ is defined, considering destructive interference for $\xi = +1$, no interference for $\xi = 0$, and constructive interference for $\xi = -1$.

Also for the monolepton selection, the trigger requirements drive the analysis thresholds. Events are selected if a well-identified and isolated electron or muon is present with $p_{\rm T} ({\rm e}) > 100 \, \mathrm{GeV}$ and $p_{\rm T} ({\rm \mu}) > 45 \, \mathrm{GeV}$, respectively. To select a monolepton event topology, requirements are imposed which ensure a large azimuthal separation between the lepton and missing momentum vector, and which check the compatibility of the transverse components of the lepton and missing momenta. After these selections, the main irreducible background comes from leptonic ${\rm W}$ decays.

\begin{figure}[htb!]
  \begin{center}
    \includegraphics[scale=0.35]{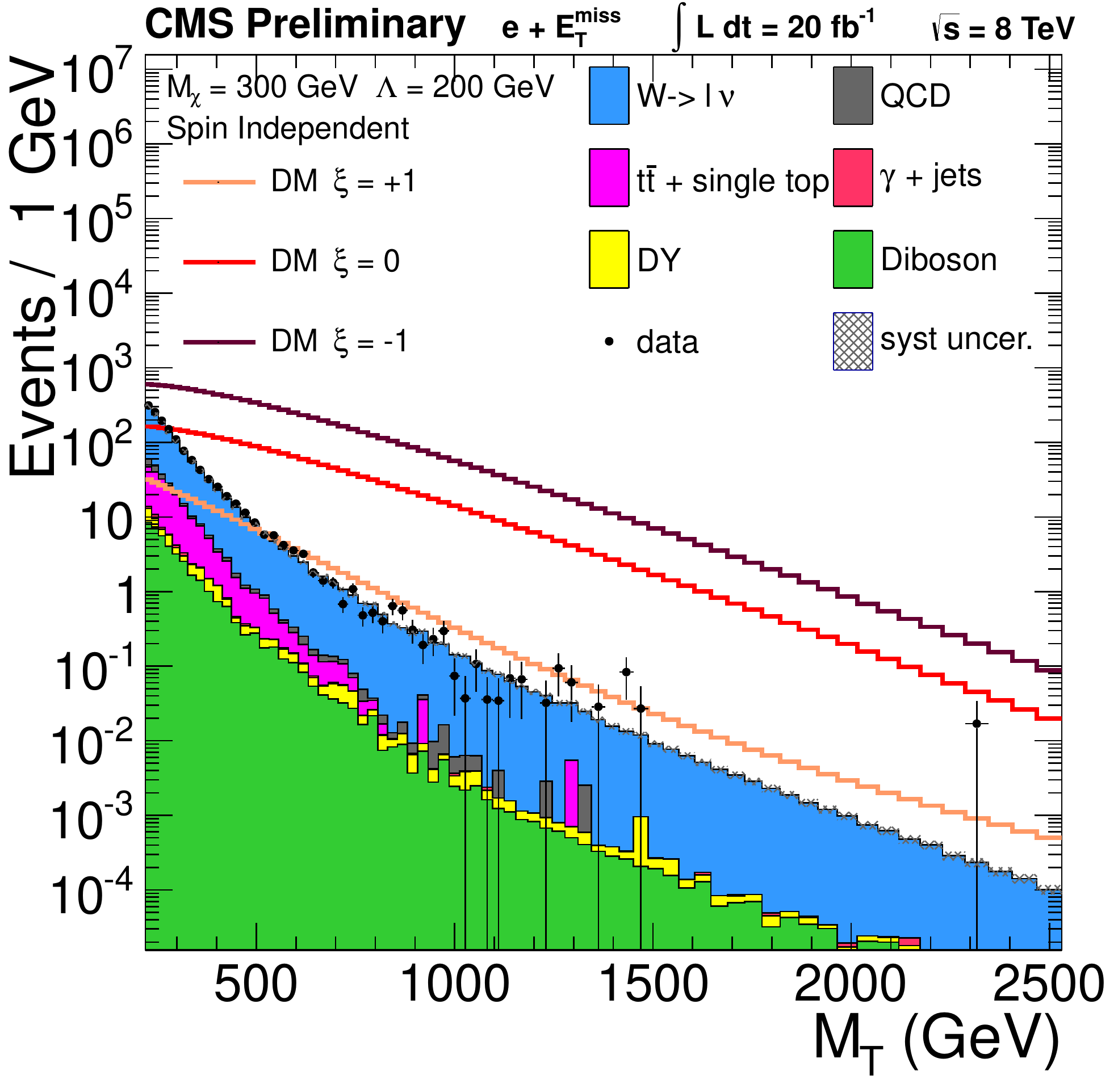}
  \end{center}
  \caption{Distribution of transverse mass after the electron selection described in the text, showing data, estimated background contributions, and signal expectation for an example model of dark matter production in the three considered cases of interference.}
  \label{fig:monoleptonele}
\end{figure}

\begin{figure}[htb!]
  \begin{center}
    \includegraphics[scale=0.35]{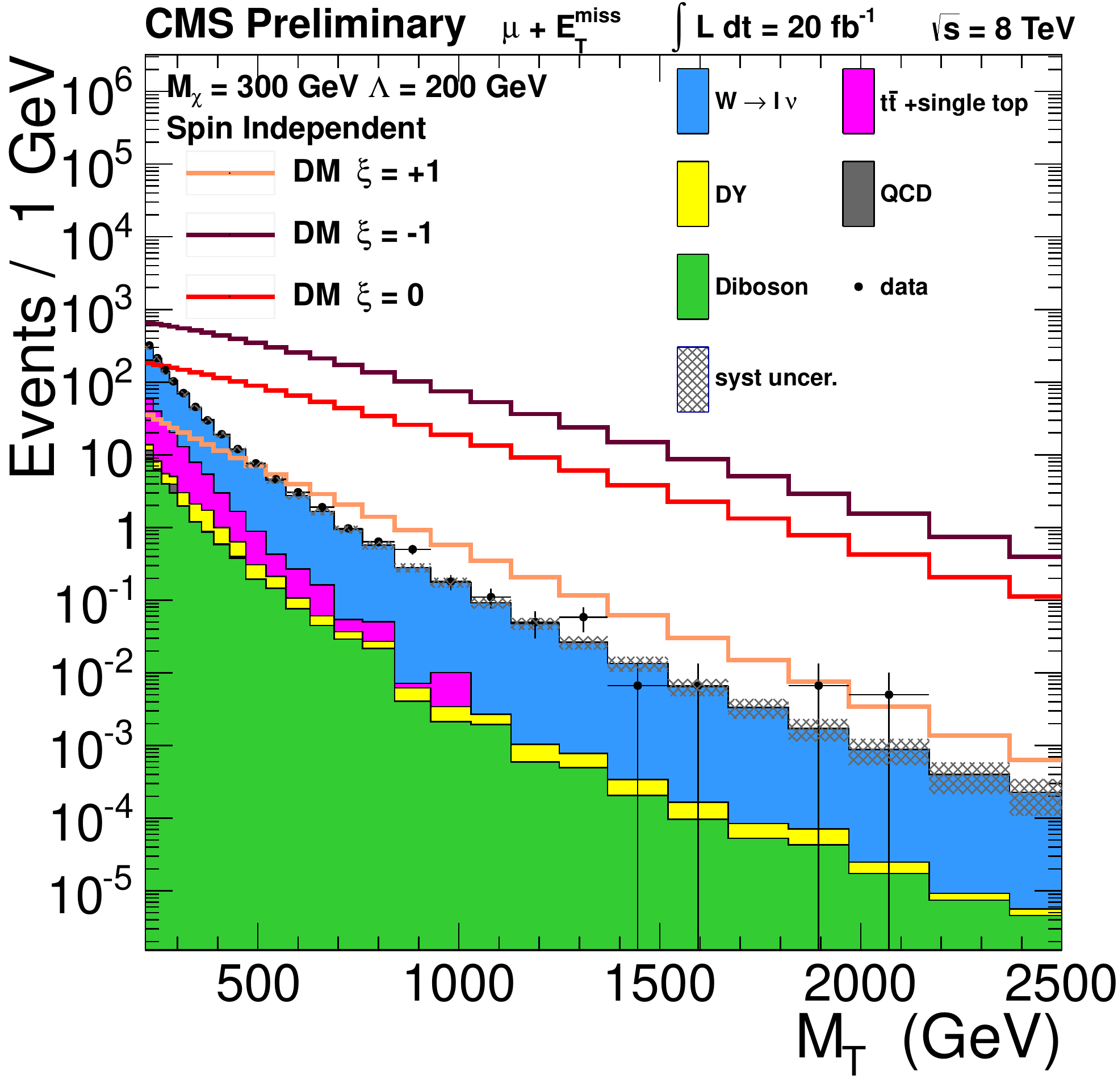}
  \end{center}
  \caption{Distribution of transverse mass after the muon selection described in the text, showing data, estimated background contributions, and signal expectation for an example model of dark matter production in the three considered cases of interference.}
  \label{fig:monoleptonmu}
\end{figure}

In Figures~\ref{fig:monoleptonele} and~\ref{fig:monoleptonmu}, the characteristic transverse mass distribution is shown, respectively for the ${\rm W} \to {\rm e}$ and  ${\rm W} \to \mu$ selection for the data, the expected background, and the three considered interference scenarios of a particular dark matter scenario.

\section{Interpretation of the monojet, monophoton and monolepton searches}

All three searches presented above find the data to be compatible with the background expectations. Hence, the analyses proceed to set limits on the possible presence of dark-matter signals in the data. Standard limits on the visible cross section of a potential new physics signal are complemented with limits on the interaction scale $\Lambda$ of the EFT approach as a function of the mass of the dark matter particle, for each considered EFT interaction operator. These limits can then in turn be translated to the plane of the dark-matter--nucleon elastic scattering cross section versus the dark matter particle mass, in which results from direct-detection experiments are usually shown.

\begin{figure}[htb!]
  \begin{center}
    \includegraphics[scale=0.35]{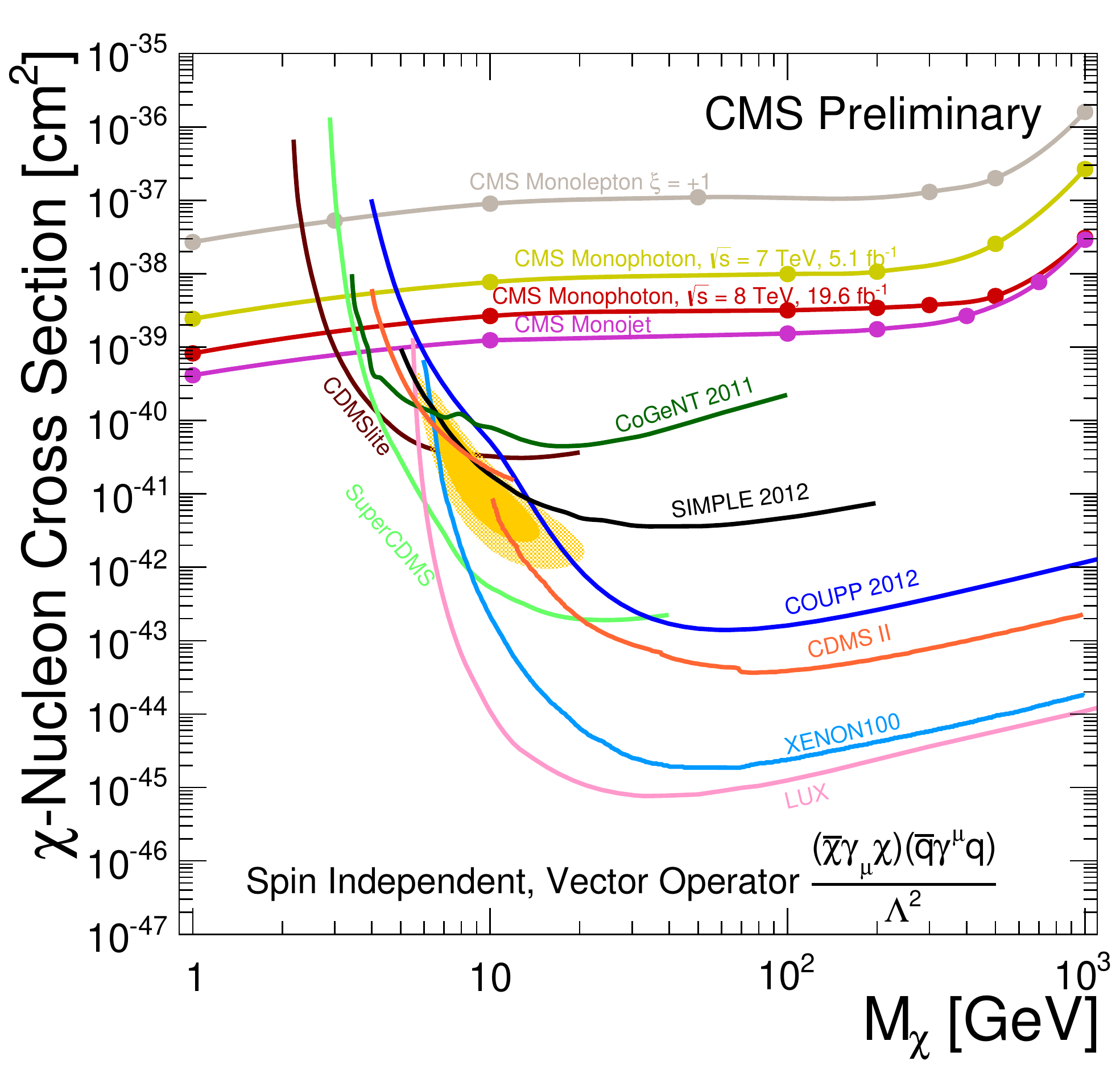}
  \end{center}
  \caption{90\%CL upper limits on the dark-matter--nucleon scattering cross section, from the monojet, monophoton, and monolepton ($\xi=+1$) searches, as a function of the dark matter mass, for spin-independent (vector operator) interactions}
  \label{fig:spinindep}
\end{figure}

\begin{figure}[htb!]
  \begin{center}
    \includegraphics[scale=0.35]{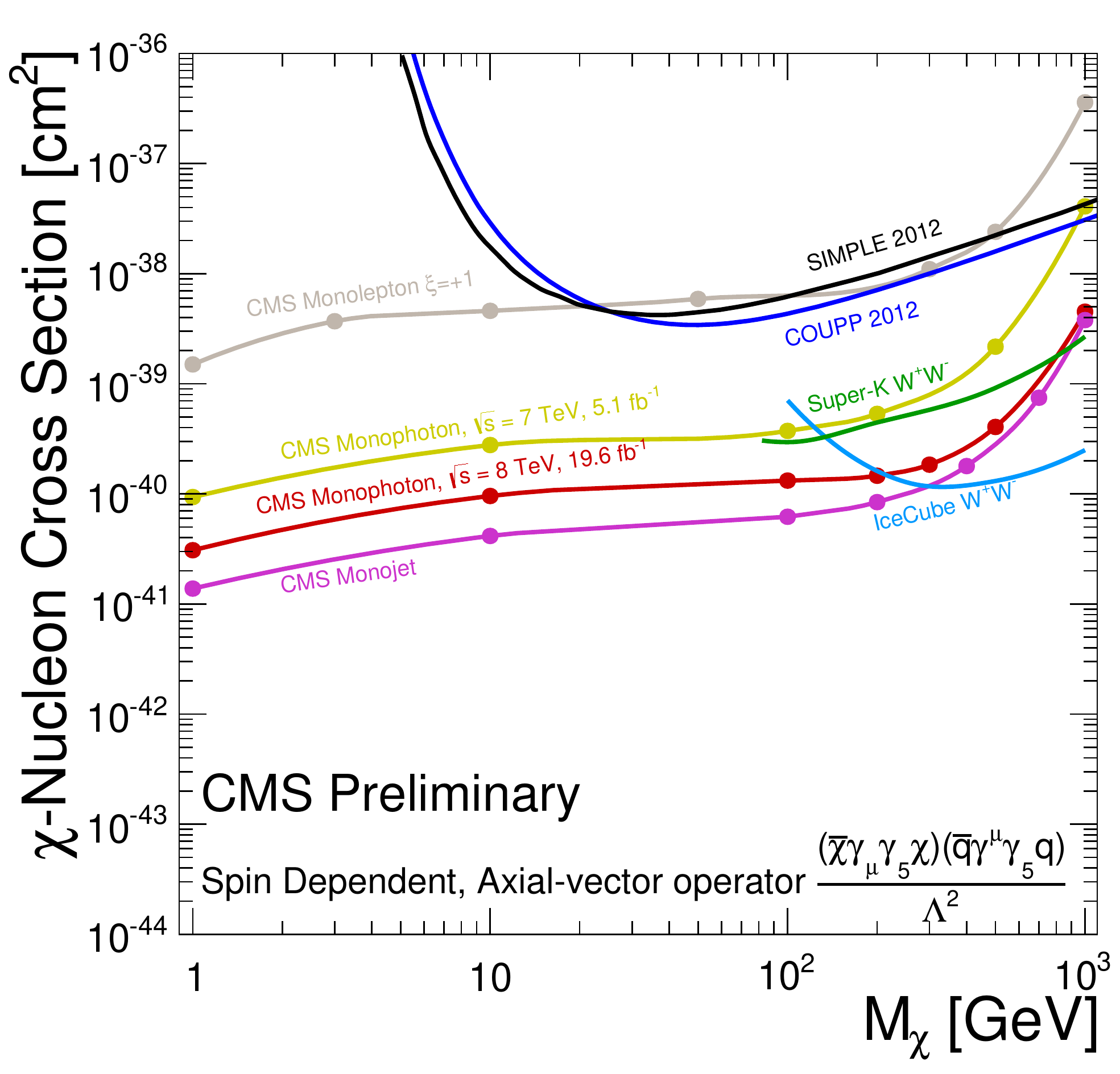}
  \end{center}
  \caption{90\%CL upper limits on the dark-matter--nucleon scattering cross section, from the monojet, monophoton, and monolepton ($\xi=+1$) searches, as a function of the dark matter mass, for spin-dependent (axial-vector operator) interactions}
  \label{fig:spindep}
\end{figure}

In Figures~\ref{fig:spinindep} and~\ref{fig:spindep}, the 90\%CL upper limits on the dark-matter--nucleon scattering cross section, from the monojet, monophoton, and monolepton ($\xi=+1$) searches, are shown as a function of the dark matter mass, for spin-independent (vector operator) and spin-dependent (axial-vector operator) interactions, respectively. Comparisons are made with results from several direct and indirect detection experiments. While it should be stressed to keep the aforementioned caveats on the interpretation of the EFT resuls in mind, a few robust observations can be made on the complementarity between the collider and direct searches. The first striking feature is the strength of the collider analyses searching for low-mass dark-matter particles. Indeed, where the recoil signals in the direct searches become too soft at low mass for efficient detection, the collider setting allows to maximize the generated missing momentum, and hence sensitivity, at zero mass. Another complementarity can be seen when comparing the two plots: the direct-detection experiments have typically reduced or no sensitivity to spin-dependent interactions, which allows the collider searches to provide complemetary coverage also at intermediate masses. At higher mass, the collider searches run out of steam because the production cross section drops -- here the indirect searches with neutrino telescopes are probing complementary ground.

The monolepton result shown in Figures~\ref{fig:spinindep} and~\ref{fig:spindep} is the most pessimistic case of destructive interference. As can be seen from Figures~\ref{fig:monoleptonele} and~\ref{fig:monoleptonmu}, the cross section may be much higher for other interference scenarios. Correspondingly, the limits on the dark-matter--nucleon scattering cross section from the monolepton search may be much stronger, even surpassing the monojet sensitivity. More details may be found in~\cite{cmsmonolepton}.

\begin{figure}[htb!]
  \begin{center}
    \includegraphics[scale=0.35]{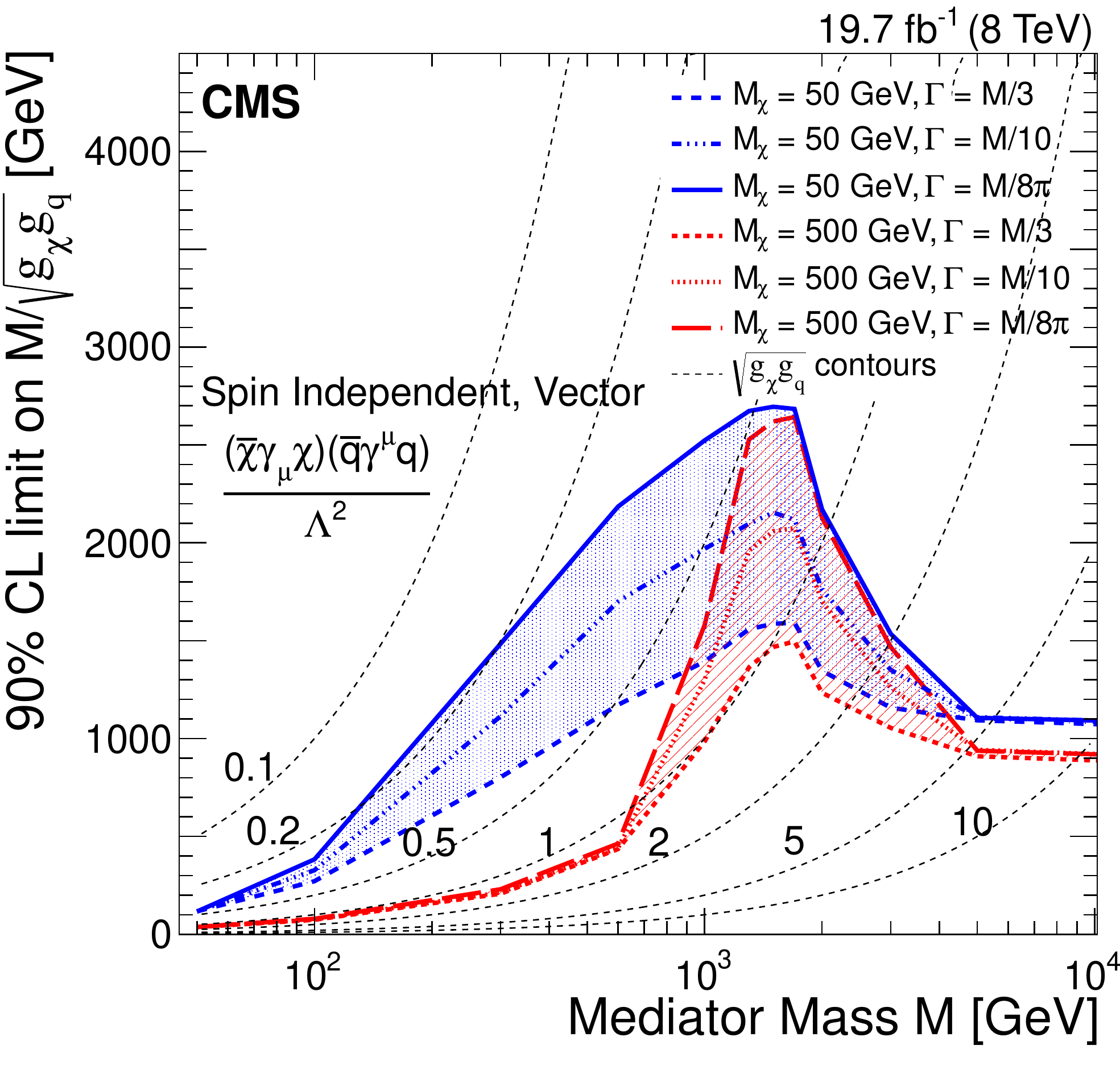}
  \end{center}
  \caption{Limits at 90\% CL on the interaction scale $\Lambda$, as a function of the mediator mass in a simplified model with an s-channel vector mediator providing the coupling between the quarrks and the dark-matter particles. Several mass and width assumptions are considered.}
  \label{fig:sms}
\end{figure}

A first effort has been pursued, in the context of the monojet analysis, to move beyond the EFT interpretation, and make the mediator explicit by means of a simplified model. In the studied case, the mediator is considered to be a vector particle. The limit on the interaction scale $\Lambda$ is calculated as a function of the mediator mass, and a range of decay width for the mediator is considered. In Figure~\ref{fig:sms}, the result of this mediator mass scan is shown. Three regimes can be discerned. At high mass, the obtained limit coincides with the EFT expectation. When decreasing the mediator mass, the mediator can go on-shell, and resonant production boosts the cross section and hence limit beyond what is naively expected from the EFT approach. For even lower mediator masses, the mediator goes off-shell again, and the limit on $\Lambda$ decreases below the naive EFT approximation, making the EFT limit too aggressive with respect to a realistic model with an explicit mediator.

\section{Searches in final states with top quarks}

Two other searches are presented for dark matter, this time leading to final states with missing energy and a single~\cite{cmsmonotop} or two top quarks~\cite{cmsditopmet}. In the case of a single top quark, referred to as a monotop final state, the dark matter particle is assumed long-lived, and couples to the top quark through flavour-changing diagrams. The second analysis, looking for two top quarks with missing energy, considers an EFT scenario where the dark-matter preferentially couples to heavy quarks, like is the case for a scalar interaction with a coupling proportional to the mass of the interacting quark.

\begin{figure}[htb!]
  \begin{center}
    \includegraphics[scale=0.35]{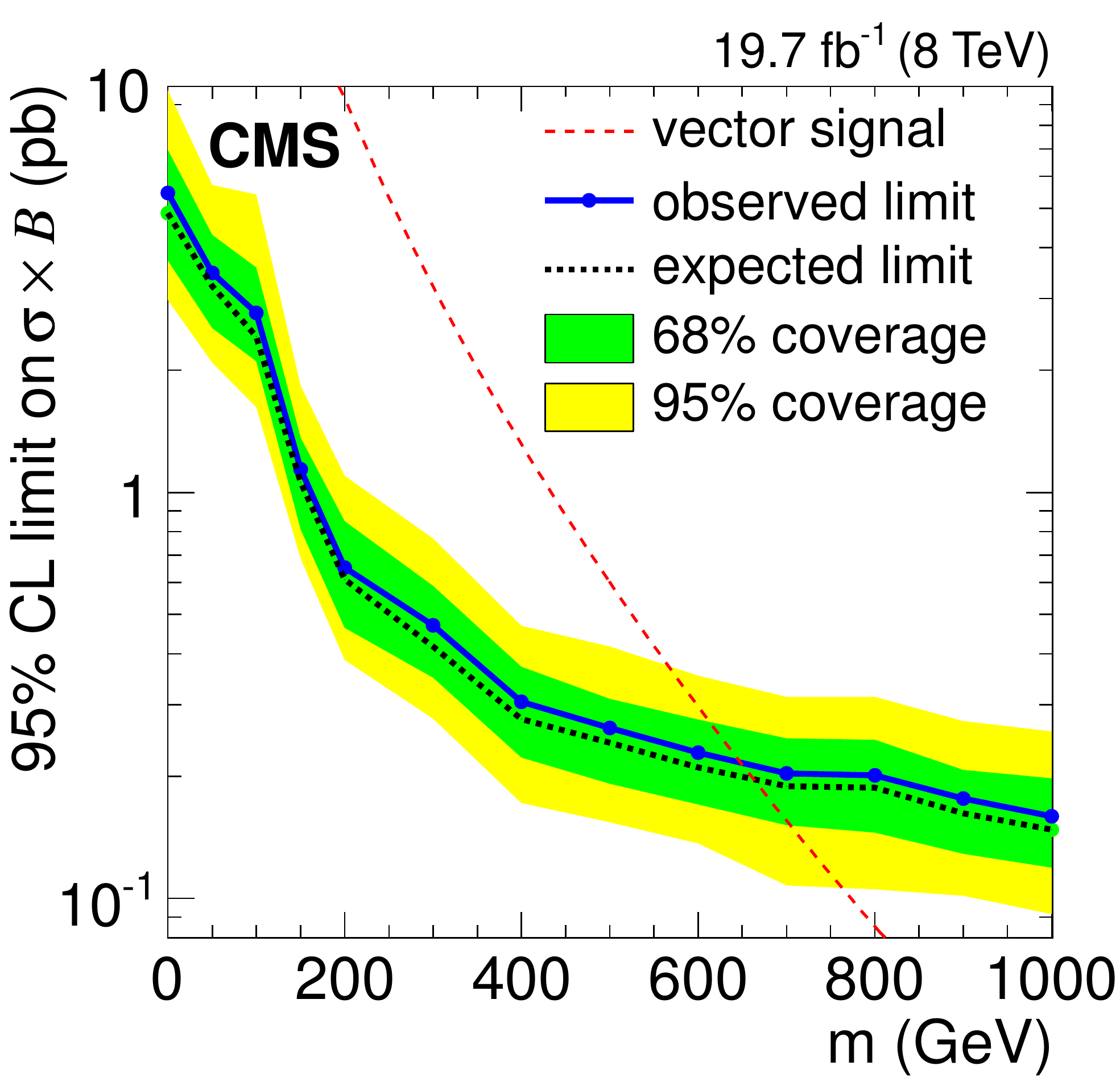}
  \end{center}
  \caption{Cross section upper limit at 95\% CL as a function of the dark-matter candidate mass for a monotop scenario with a vector dark-matter particle. The cross section of the considered model is shown as the dashed red line. Masses below $650\,\mathrm{GeV}$ are excluded at 95\% CL.}
  \label{fig:monotop}
\end{figure}

The selection for the monotop search selects hadronic final states by requiring a large missing momentum, $E_{\rm T}^{\rm miss} > 350 \, \mathrm{GeV}$, and 3 jets, of which one is identified as a ${\rm b}$ quark. Additionally, an electron and muon veto is applied to suppress backgrounnds with genuine missing energy from the neutrino in leptonic ${\rm W}$ decays. This selection leaves ${\rm t}\bar{\rm t}$ and ${\rm Z}$+jets as the main backgrounds. The total background expectation is $28 \pm 16$ events, while $30$ events are observed in the data. In absence of an excess, limits were determined on the possible presence of a scalar and vector dark-matter particle in this monotop scenario. In Figure~\ref{fig:monotop}, the 95\% CL upper limit on the cross section is shown as a function of the mass of the dark-matter candidate, in the case it is a vector particle. This is compared with the production cross section of the considered model, leading to this scenario being excluded at 95\% CL for masses below $650\,\mathrm{GeV}$. A scalar dark-matter candidate is similarly excluded for masses below $330\,\mathrm{GeV}$.

\begin{figure}[htb!]
  \begin{center}
    \includegraphics[scale=0.35]{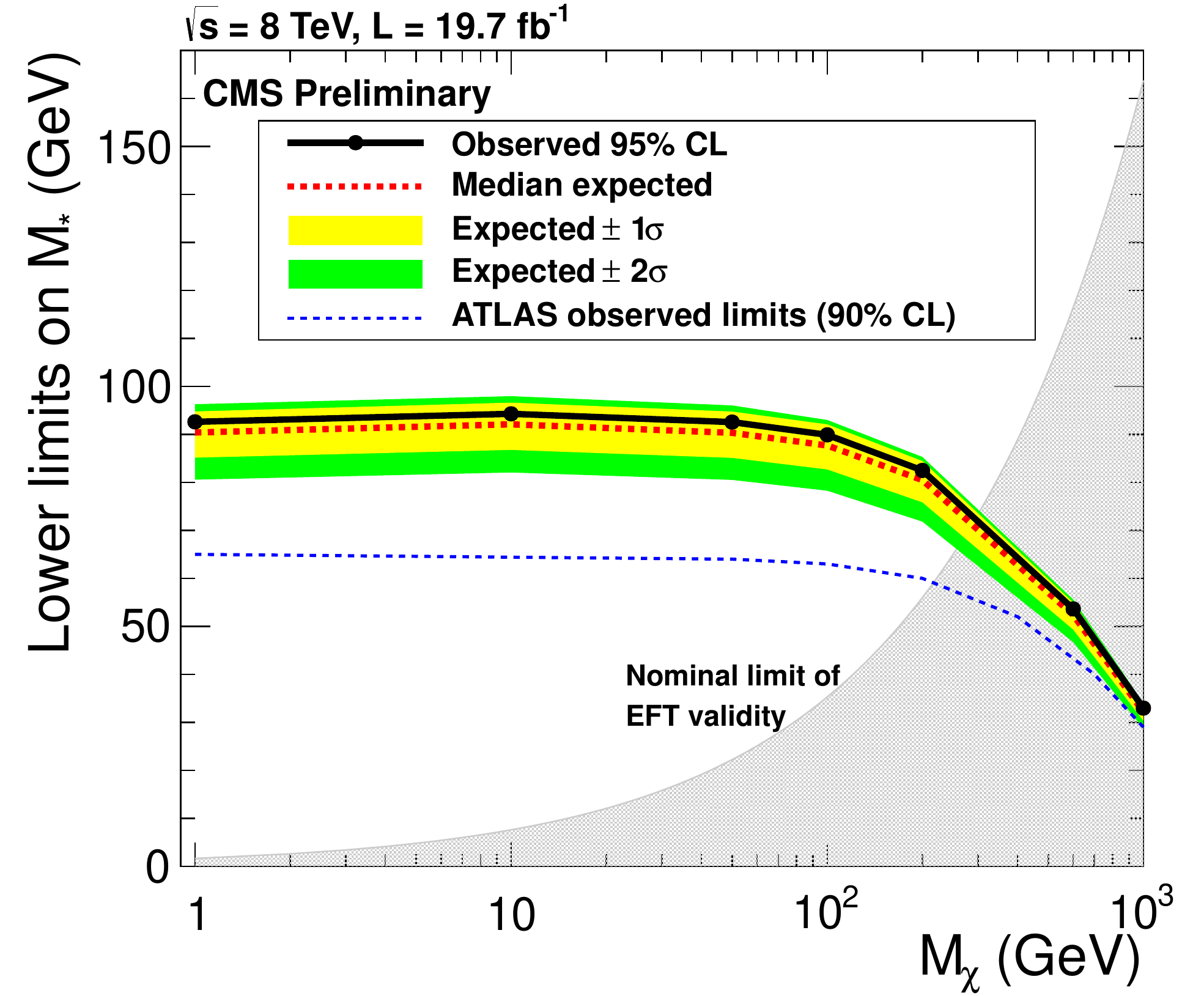}
  \end{center}
  \caption{Lower limits at 95\% CL on the scale of the scalar EFT interaction considered, as a function of the mass of the dark-matter particle.}
  \label{fig:ditopmet}
\end{figure}

The selection for the ${\rm t}\bar{\rm t}$ + $E_{\rm T}^{\rm miss}$ final state aims for the dilepton decay channel. Two well-identified electrons or muons are required, along with two or more jets, and $E_{\rm T}^{\rm miss} > 320\,\mathrm{GeV}$. Further cuts are applied on the opening angle between the leptons, and on the scalar sums of the transverse momenta of the leptons on the one hand, and the jets on the other. The background remaining after these selection cuts is dominated by top quarks, with a non-negligible contribution from diboson and Drell-Yan events. The total background is estimated to be $1.9 \pm 0.7$ events, while in data 1 event is observed to pass the selection. With background expectation and data being compatible, lower limits are set on the interaction scale in the described EFT context. In Figure~\ref{fig:ditopmet}, these limits are shown as a function of the mass of the dark-matter particle. Expressed as cross section upper limits, the considered EFT scenarios are excluded at 95\% CL for cross sections larger than $0.24 \, \mathrm{pb}$ and $0.09 \, \mathrm{pb}$, for dark matter masses of $50 \, \mathrm{GeV}$ and $1000 \, \mathrm{GeV}$, respectively.

\nocite{*}
\bibliographystyle{elsarticle-num}
\bibliography{lowette}







\end{document}